\documentclass[aps,pra,twocolumn,showpacs,superscriptaddress,longbibliography]{revtex4-1}
\usepackage{makeidx}
\usepackage[utf8]{inputenc}
\usepackage{graphicx}
\usepackage{amsmath}
\usepackage{amsfonts}
\usepackage{mathrsfs}
\usepackage{graphicx}
\usepackage{braket}
\usepackage[subrefformat=parens,labelformat=parens,caption=false]{subfig}
\usepackage{dcolumn}
\usepackage{bm}
\usepackage{bbm}
\usepackage{color}
\usepackage[dvipsnames]{xcolor}
\usepackage{esint}
\usepackage{lineno}
\usepackage{float}
\usepackage{verbatim}
\usepackage[breaklinks,
            colorlinks,
            urlcolor=blue,
            linkcolor=blue,
            anchorcolor=blue,
            citecolor=blue]{hyperref}

\usepackage{color}
\definecolor{KB}{rgb}{0.4,0.3,0.9}

\usepackage{soul}

\begin{document}
\title{Optimal Probes for Global Quantum Thermometry}

\author{Wai-Keong Mok}
\affiliation{Centre for Quantum Technologies, National University of Singapore, 3 Science Drive 2, Singapore 117543}
\author{Kishor Bharti}
\affiliation{Centre for Quantum Technologies, National University of Singapore, 3 Science Drive 2, Singapore 117543}
\author{Leong-Chuan Kwek}
\affiliation{Centre for Quantum Technologies, National University of Singapore, 3 Science Drive 2, Singapore 117543}
\affiliation{MajuLab, CNRS-UNS-NUS-NTU International Joint Research Unit, UMI 3654, Singapore}
\affiliation{National Institute of Education and Institute of Advanced Studies,
Nanyang Technological University, 1 Nanyang Walk, Singapore 637616}
\affiliation{School of Electrical and Electronic Engineering Block S2.1, 50 Nanyang Avenue,
Singapore 639798 }
\author{Abolfazl Bayat}
\affiliation{Institute of Fundamental and Frontier Sciences,
University of Electronic Science and Technology of China, Chengdu 610051, China}

\begin{abstract}
Quantum thermodynamics has emerged as a separate sub-discipline, revising the concepts and laws of thermodynamics, at the quantum scale. In particular, there has been a disruptive shift in the way thermometry, and thermometers are perceived and designed. Currently, we face two major challenges in quantum thermometry. First, all of the existing optimally precise temperature probes are local, meaning their operation is optimal only for a narrow range of temperatures. Second, aforesaid optimal local probes mandate complex energy spectrum with immense degeneracy, rendering them impractical. Here, we address these challenges by formalizing the notion of global thermometry leading to the development of optimal temperature sensors over a wide range of temperatures. We observe the emergence of different phases for such optimal probes as the temperature interval is increased. In addition, we show how the best approximation of optimal global probes can be realized in spin chains, implementable in ion traps and quantum dots.

\end{abstract}

\date{\today}
\maketitle

\section{Introduction} 
\begin{figure*}
\subfloat{%
  \includegraphics[width=0\linewidth]{example-image-a}%
  \label{fig:scheme}
}
\subfloat{%
  \includegraphics[width=0\linewidth]{example-image-a}%
  \label{fig:qudit16_bifurcation}
}
\subfloat{%
  \includegraphics[width=0\linewidth]{example-image-a}%
  \label{fig:qudit64_bifurcation}
}
\subfloat{%
  \includegraphics[width=0\linewidth]{example-image-a}%
  \label{fig:tau_scaling}
}
\subfloat{%
  \includegraphics[width=\linewidth]{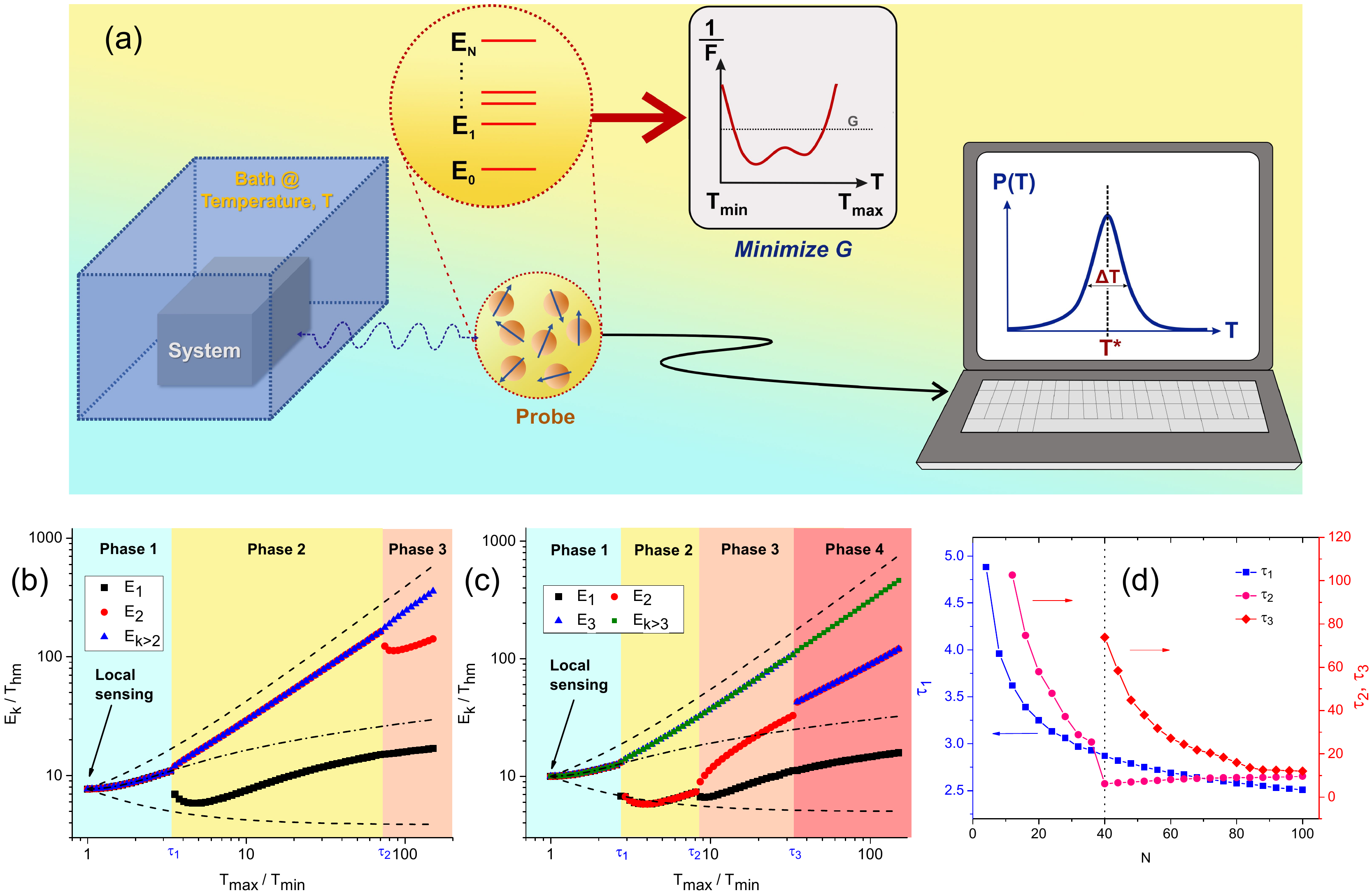}%
}\hfill
	\caption{\textbf{Global quantum thermometry.} (a) Schematic of our proposal. A probe weakly interacts with the system and thus reaches an equilibrium at the system temperature $T \in [T_{\text{min}},T_{\text{max}}]$. Energy measurements on the probe yield a probability distribution of temperatures with an estimated temperature $T^*$ and variance $(\Delta T)^2$. The global thermometry measure $G$ defined in Eq. (\ref{eq:global_measure}) is a property of the probe energy spectrum, and is minimized to achieve the optimal probe. The optimal energies against temperature ratio $T_\text{max}/T_\text{min}$ for a general $N$-level system is obtained for: (a) $N = 16$, (b) $N = 64$. The energies are scaled with the characteristic temperature $T_{hm}$, where $T_{hm}$ is the harmonic mean of $T_\text{min}$ and $T_\text{max}$. The leftmost data point corresponds to the case of the locally optimal thermometer. The dashed lines indicate the approximate bounds for the energies (explained in main text), while the dash-dotted line shows the optimal energy gap for an effective two-level system. (d) Critical ratios $\tau_k$ against $N$.}
	\label{fig:qudit_bifurcation}
\end{figure*}

Recent advancements in quantum technologies have pushed the boundaries of thermodynamics into new territories, where small objects are cooled to ultra-low temperatures~\cite{bloch2008many, giazotto2006opportunities}. This has led to the development of a fast-growing field of quantum thermodynamics~\cite{vinjanampathy2016quantum,mehboudi2019thermometry,binder2018thermodynamics,RevModPhys.83.771}. From a fundamental perspective,  this provides new definitions for some of the old concepts, such as work~\cite{PhysRevLett.118.070601,talkner2016aspects} and heat~\cite{esposito2010entropy}, a new formulation for the laws of thermodynamics~\cite{brandao2015second,uzdin2018global,kolavr2012quantum,levy2012quantum} and scaling relations~\cite{bayat2016nonequilibrium}. From a practical viewpoint, however, measuring the thermodynamic quantities at the quantum level 
mandates unprecedented precision~\cite{degen2017quantum} and advanced refrigeration~\cite{he2002quantum,timofeev2009electronic,timofeev2009electronic,mohammady2018low}. 
Temperature, defined through the zeroth law of thermodynamics, is one of the key thermodynamical parameters, with its precise measurement having numerous applications from our daily life to almost any quantum experiment. Thermometry has been explored on various experimental platforms \cite{mehboudi2019thermometry}, including nitrogen-vacancy (NV) centres in diamonds \cite{toyli2013fluorescence,kucsko2013nanometre}, optical nanofibres \cite{grover2015photon} and nano-photonic cavities coupled to nano-optomechanical resonators \cite{purdy2017quantum}.

At low temperatures, precise temperature measurement is extremely challenging \cite{stace2010quantum,potts2019fundamental,brunelli2011qubit,brunelli2012qubit,sabin2014impurities,mehboudi2015thermometry,guo2015improved,hofer2017quantum,de2017estimating,campbell2017global,campbell2018precision,plodzien2018few,sone2018quantifying}. To measure the temperature of a quantum system, one brings in a probe, i.e. thermometer, and interact it with the system. Then, the procedure of thermometry can be pursued in two ways: (i) non-equilibrium dynamics in which the temperature is extracted as a parameter from the state of the probe before its thermalization~\cite{de2017estimating,brunelli2011qubit,montenegro2020mechanical,brunelli2012qubit,feyles2019dynamical,gebbia2020two,mancino2020non,mitchison2020situ,mukherjee2019enhanced,seah2019collisional}; and (ii) equilibrium approach in which the probe is measured after reaching an equilibrium with the system and, thus, is described by a thermal state with the same temperature as the system~\cite{stace2010quantum,Louis2015,paris2015achieving,campbell2018precision,correa2017enhancement,de2016local,razavian2019quantum}. In the first approach, the optimal measurement, which minimizes the uncertainty of the estimation through saturating the Cram\'{e}r-Rao inequality~\cite{Cramer1999mathematical}, generally depends on the unknown temperature of the system which makes its achievement challenging in practice. In the second approach, which is the focus of this paper, the energy measurement is known to be the optimal choice which saturates the quantum Cram\'{e}r-Rao bound~\cite{stace2010quantum} independent of temperature. Nonetheless, not all probes provide the same precision in equilibrium thermometry. Therefore, one may wonder what an optimal probe for equilibrium quantum thermometry is.    

In Ref.~\cite{Louis2015}, the above question has been answered analytically for a general $N$-level quantum probe. The result shows that the probe has to be an effective two-level system with a unique ground state and $N-1$ degenerate excited states. Unlike the optimal measurement basis, the energy gap of the optimal probe is temperature-dependent. This means that the design of an optimal probe needs prior information about the temperature of the system. In other words, the probe can only be used for local thermometry in which the temperature is roughly known, and the objective of thermometry is just to measure it more precisely. Moreover, this optimal probe, with enormous degeneracy in its excited state, is very difficult to be realized in practice. 
Based on these, there are several natural questions that one can ask: (i) How can we design a probe which operates optimally over a wide range of temperatures? (ii) What does the optimal probe look like if one takes the practical constraints into account?  

In this paper, we address all of the aforementioned questions. We first quantitatively develop the concept of global thermometry over an arbitrary temperature interval and introduce the average variance as the quantity which has to be minimized to get the optimal probe. Our notion of global thermometry is distinct from existing works in literature, where the word ``global'' refers to system size \cite{campbell2017global} and not temperature range. Using our approach, we can determine the optimal probe over a wide range of temperature intervals. Interestingly, the structure of the optimal probe goes through different phases, each with its own energy structure, as the temperature interval increases. We then focus on spin chain probes with various constraints to get the structure of the optimal probes considering practical limitations which one may face in realistic scenarios. Our results show that for the non-uniform XYZ Hamiltonians, the Ising spin chain provides the optimal probe. Interestingly, by considering the non-uniform Heisenberg Hamiltonians, the dimer spin chains become optimal. Our results pave the way for global thermometry in various physical systems, including ion traps and quantum dot arrays.

\section{Local quantum thermometry}
\label{sec:local_thermo}

In this section, we review the concept of local quantum thermometry at equilibrium and its optimal probes. Let us consider a thermometer that weakly interacts with a system at some temperature $T$ (see scheme in Fig. \subref*{fig:scheme}). At thermal equilibrium, we can assume that the probe thermalizes at the system temperature $T$, and is thus described by the Gibbs state $\rho_{th} = \exp{(-\beta H_p)}/\mathcal{Z}$, where $\beta \equiv 1/k_B T$ is the inverse temperature (we set $k_B = 1$ henceforth), $H_p$ is the probe Hamiltonian with $N$ eigenvalues and $\mathcal{Z}$ is the partition function. For $\mathcal{M}$ measurement samples, the variance in the estimation of temperature (see Fig. \subref*{fig:scheme}) satisfies the Cram\'{e}r-Rao inequality:
	\begin{equation}
(\Delta T)^2 \geq \frac{1}{\mathcal{M} \mathcal{F}_{th}}
\label{Cramer-Rao-ineq}
	\end{equation}
where $\mathcal{F}_{th}$ is the thermal quantum Fisher information (QFI) at equilibrium, which can be related to the heat capacity and hence the energy variance of the probe~\cite{mehboudi2019thermometry}:
	\begin{equation}
\mathcal{F}_{th} = \frac{ (\Delta H_p)^2 } {T^4} = \frac{ \braket{H_p^2} - \braket{H_p}^2 }{T^4}
	\end{equation}


It is worth emphasizing that the Cram\'{e}r-Rao bound is saturated for all temperatures if one measures the probe in the energy basis~\cite{stace2010quantum}, for which $\rho_{th}$ is diagonal. Thus, in equilibrium thermometry, the only way to increase precision is by increasing the thermal QFI through proper engineering of the probe. However, $\mathcal{F}_{th}$ depends on the actual energy spectrum of the probe. An interesting question then arises: what is the optimal Hamiltonian spectrum for a $N$-level system that maximizes the thermal QFI? This question was answered in a previous work by Correa \textit{et al.} \cite{Louis2015}, where they proved that the optimal energy spectrum is an effective two-level system with a single ground state and a $(N-1)$-fold degenerate excited state. They also showed that the energy gap $\epsilon$ between the excited state and ground state is the solution of a transcendental equation \cite{Louis2015}:
	\begin{equation}
e^x = (N-1) \frac{x+2}{x-2}
	\label{eq:local_opt}
	\end{equation}
where $x \equiv \epsilon / T$ is the dimensionless energy gap. 

Nonetheless, this optimal thermometry has a serious limitation as from Eq.~(\ref{eq:local_opt}), one can only obtain the ratio $\epsilon/T$. Hence, for a fixed number of energy levels $N$, the optimal energy gap $\epsilon$ scales linearly with the temperature $T$. However, $T$ is generally unknown to us. Thus, the probe cannot be optimized unless the temperature is already known within a narrow range, which limits the usefulness of the probe. Due to this limitation, this scheme is known as \textit{local} thermometry.

\section{Global quantum thermometry}
\label{sec:global_thermo}

In order to go beyond the limitations of local thermometry, one has to first define the paradigm of global thermometry in which the probe can operate optimally over a wide range of temperatures. We define the average variance $G(T_{\text{max}}, T_{\text{min}})$ over some given temperature range $[T_\text{min}, T_\text{max}]$ as 
\begin{equation}
	\begin{split}
G(T_{\text{max}}, T_{\text{min}}) &\equiv \frac{1}{T_{\text{max}}-T_{\text{min}}} \int_{T_{\text{min}}}^{T_{\text{max}}} \frac{dT}{\mathcal{F}_{th} (T)} 
    \end{split}
    \label{eq:global_measure}
\end{equation}
where the right hand side is proportional to the average of Cram\'{e}r-Rao bound in Eq.~(\ref{Cramer-Rao-ineq}). Unlike the local thermometry scheme, in which the QFI is maximized, in the global thermometry one has to minimize the average variance $G(T_{\text{max}}, T_{\text{min}})$. In the limit of $T_{\text{max}}/T_{\text{min}} \to 1$, $G \to 1/\mathcal{F}_{th}$, and the minimization of $G$ becomes equivalent to maximizing $\mathcal{F}_{th}$. This recovers the special case of local thermometry. The temperature estimation can also be done adaptively in the sense that once the temperature is estimated for a given range then one can then use a secondary probe optimized for a tighter range to get a better estimate. This procedure can be repeated until the desired precision is achieved.

%
For a given probe of size $N$, one has to optimize the $N-1$ energy levels to minimize $G(T_{\text{max}}, T_{\text{min}})$, setting the ground state energy $E_0=0$ as a reference point. The aforementioned integral in Eq.~(\ref{eq:global_measure}) is very difficult to compute analytically (see Appendix \ref{appendixA} for some analytical approximations), so we use numerical optimization to minimize $G$ over all the $N-1$ independent energy levels. We note that for large system sizes, especially for the spin chains, the optimization routine becomes infeasible due to the computational costs from diagonalizing the Hamiltonian and the large number of iterations required to ensure a global optimum. In such cases, we combine local optimization with transfer learning to speed up the convergence to optimality (see Sec. \ref{sec:methods} for more details).

Figs. \subref*{fig:qudit16_bifurcation} and \subref*{fig:qudit64_bifurcation} show the optimized energy spectrum for $N = 16$ and $N = 64$ energy levels, for a temperature range of up to $T_{\text{max}}/T_{\text{min}} = 150$. We use the harmonic mean temperature $T_{hm} \equiv (1/T_{\text{max}} + 1/T_{\text{min}})^{-1}$ as the characteristic temperature, and plot the energy levels normalized by $T_{hm}$ against the temperature ratio $T_{\text{max}} / T_{\text{min}}$. Since the ground state is fixed to be $E_0 = 0$, we only plot the energy levels of the excited states, where $E_k$ is the energy of the $k^{\text{th}}$ excited state. In the limiting case where $T_{\text{max}} / T_{\text{min}} = 1$, the situation reduces to the local thermometry case studied in Ref. \cite{Louis2015}. Unsurprisingly, numerical optimization gives the energy gap $E_1 = E_2 = \ldots E_N \approx 7.708 \,  T_{hm}$, which is consistent with Eq. (\ref{eq:local_opt}). This exactly describes the effective two-level solution obtained in Ref. \cite{Louis2015}. As the ratio $T_{\text{max}}/T_{\text{min}}$ is increased further, the optimal solution is again an effective two-level system with a $(N-1)$-fold degenerate excited state. We call this regime of the optimal energy spectrum, namely an effective two level system, Phase 1. The energy gap here, of course, cannot be described by Eq. (\ref{eq:local_opt}), as both $T_{\text{min}}$ and $T_{\text{max}}$ have to be taken into account. The corrections to Eq. (\ref{eq:local_opt}) for the regime $T_{\text{max}} /T_{\text{min}}$ near unity can be found in Appendix \ref{appendixA}.

The upper and lower dashed lines in Figs. \subref*{fig:qudit16_bifurcation} and \subref*{fig:qudit64_bifurcation} denotes the optimal energy gap when the probe is optimized for only $T_{\text{max}}$ or $T_{\text{min}}$ respectively. This gives a nice physical intuition for the optimal energy gap, as a compromise between the locally optimal probes at $T_{\text{max}}$ and $T_{\text{min}}$. A simple estimate of the energy gap is the average energy between the upper and lower bounds, which is equivalent to optimizing the probe at the average temperature $(T_{\text{max}} + T_{\text{min}})/2$. This is fairly accurate, within 3 \% of the actual energy gap in Phase 1 for $N = 16$. 

As the temperature ratio $T_{\text{max}} / T_{\text{min}}$ is increased beyond some critical ratio $\tau_1$, the optimal energy spectrum is no longer an effective two-level system. Instead, the energy levels remarkably bifurcate into two branches creating a new regime of energy spectrum, namely Phase 2. The optimal probe in Phase 2 is an effective three-level system. The dashed-dotted lines in 
Figs. \subref*{fig:qudit16_bifurcation} and \subref*{fig:qudit64_bifurcation} show the optimization results when the probe is constrained to an effective two-level system, and, can be thought of as an extrapolation of the Phase 1 solutions. 

The bifurcation of energy levels can be physically interpreted as the case where the optimal probe has to be sensitive to both ends of the temperature range. When the range is small (i.e. $T_{\text{max}} / T_{\text{min}}$ below the critical ratio), optimizing the probe at a temperature near the mean gives a sufficiently high $F_{th}$ at both ends of the range. The optimal solution is the highly-degenerate two-level system which has been previously found to be the most sensitive probe for local thermometry \cite{Louis2015}. This argument is obviously not true in the limit where the temperature range is large (i.e. $T_{\text{max}} / T_{\text{min}}$ above the critical ratio), and the energy gaps have to be split up to produce an effective probe over the entire temperature range. The above intuition is supported by the dashed lines, corresponding to the optimized local probes at $T_{\text{min}}$ and $T_{\text{max}}$ respectively, which provide approximate bounds for the actual energy levels. 

The huge discrepancy between the degeneracies of $E_1$ and $E_2$ (1 vs $N-2$) is consistent with previous findings~\cite{campbell2018precision}. This is because the population of the higher excited states are exponentially suppressed, which in turn demands a significantly higher degeneracy in order to achieve a reasonable sensitivity at higher temperatures. As the temperature ratio is increased further, we get further bifurcations of the energy levels in Phases 3 and 4 at the critical ratios $\tau_2$ and $\tau_3$ respectively in Fig. \subref*{fig:qudit64_bifurcation}, leading to an effective four-level system. The difference in the bifurcation diagram can be intuitively understood: for a larger N, more energy levels can be split up from the upper branch to `support' the lower temperatures without sacrificing too much sensitivity at higher temperatures. The critical ratios are plotted against $N$ in Fig. \subref*{fig:tau_scaling}. An interesting observation is that the critical ratio $\tau_2$ decreases sharply at $N = 40$, with the emergence of Phase 4 as indicated by the presence of $\tau_3$.


In short, we have discovered new phases of optimal solutions for global quantum thermometry, when the previously found effective two-level system is no longer ideal. This allows us to delineate global thermometry into various regimes. We also emphasize that our observations for the emergence of various phases for the optimal global probe are universal. In other words, it does not depend on the specific $T_{\text{min}}$ and $T_{\text{max}}$, and only depends on the ratio $T_{\text{max}}/T_{\text{min}}$ and the harmonic mean $T_{hm}$.

\section{Realization of optimal probes}

Before we study the optimization of various practical spin chain models, we first demonstrate that the optimal energy spectrum for Phase 1 thermometry, the effective two-level system, can be obtained from a generalized version of the classical (or longitudinal-field) Ising model. Consider the Hamiltonian for $n$ spin-$\frac{1}{2}$ particles $(\text{i.e. } N = 2^n)$:
	\begin{equation}
H = \sum_{k=1}^{n} H_k
    \label{eq:generalized_ising}
	\end{equation}
where 
	\begin{equation}
H_k = \sum_{\braket{i_1, i_2, \ldots i_k}} J_k^{i_1,i_2,\ldots,i_k} \sigma_z^{i_1} \sigma_z^{i_2} \ldots \sigma_z^{i_k}
	\end{equation}
is the $k$-local Hamiltonian with coupling parameters $J_k$ (omitting the spin indices), and $\sigma_z^{i}$ is the Pauli Z operator for the $i^{\text{th}}$ spin. For example, $H_1$ can be interpreted as the interaction of individual spins with their respective external magnetic fields $J_1$, while $H_2$ is the usual $ZZ$-interaction in the Ising model. Diagrammatically, this Hamiltonian can be depicted by the union of complete $k$-uniform hypergraphs ($k = 1,2,\ldots,N$) where the vertices denote the spins and the the hyperedges denote the couplings between them. 

To simplify the model, we can set the magnitudes of all the $J_k$ to be the same, i.e. $|J_k| = J$, $k = 1, 2, \ldots n$. By choosing the signs appropriately, the energy spectrum can be made such that there is one ground state and $(N - 1)$ degenerate excited states, with an energy gap of $N J$. Thus, by knowing the theoretically optimal energy gap $\epsilon(T_{\text{min}}, T_{\text{max}})$, one can tune $J = \epsilon/N$ to obtain the optimal probe for the Phase 1 thermometry. Interestingly, our numerical evidence suggests that the optimal energy structure in other phases can also be obtained with this generalized model (more details in Appendix \ref{appendixB}).

Although one could achieve the theoretical limits of equilibrium quantum thermometry with the classical Hamiltonian in Eq.~(\ref{eq:generalized_ising}), this generalized model is unrealistic since it requires full control of arbitrary $k$-body interactions. This is clearly not feasible for a reasonably large $n$. In the following subsection, we discuss the optimal global thermometry in realistic spin chain setups with much fewer control parameters.

\section{Spin chain probes}

As mentioned above, the optimal probes for global thermometry demand special form of energy structures, which are difficult to engineer with physical systems. Moreover, even if such spectra can be engineered, it might be challenging to measure the probe in the energy basis which is required to saturate the Cram\'{e}r-Rao bound. Hence, there is a need to design practical probes with a high thermometric performance (compared to the theoretical limit). In this section, instead of optimizing over all the $(N-1)$ independent energy levels, we first start with a spin chain system and minimize $G$ over the spin chain parameters which can be controlled experimentally.

Restricting ourselves to 1D spin chains with only nearest-neighbour (two-body) interactions and periodic boundary conditions, we can write down a non-uniform Heisenberg XYZ Hamiltonian:
	\begin{equation}
	\begin{split}
H_{\text{XYZ}} &= \sum_{i=1}^{n} (J_x^{i} \sigma_x^{i} \sigma_x^{i+1} + J_y^{i} \sigma_y^{i} \sigma_y^{i+1} + J_z^{i} \sigma_z^{i} \sigma_z^{i+1}) \\
&+ \sum_{i=1}^{n} (h_x^{i} \sigma_x^{i} + h_y^{i} \sigma_y^{i} + h_z^{i} \sigma_z^{i})
	\end{split}
	\end{equation}
with the presence of local external magnetic fields in all three directions. The couplings and fields are also allowed to vary independently throughout the spin chain. A further constraint $J_x^{i} = J_y^{i} = J_z^{i} = J^{i}$ can be added, which gives the non-uniform Heisenberg XXX Hamiltonian:
	\begin{equation}
	\begin{split}
H_{\text{XXX}} &= \sum_{i=1}^{n} J^{i} (\sigma_x^{i} \sigma_x^{i+1} + \sigma_y^{i} \sigma_y^{i+1} + \sigma_z^{i} \sigma_z^{i+1}) \\
&+ \sum_{i=1}^{n} (h_x^{i} \sigma_x^{i} + h_y^{i} \sigma_y^{i} + h_z^{i} \sigma_z^{i})
	\end{split}
	\end{equation}
In practice, it is difficult to engineer exchange couplings with varying signs. Therefore, we impose the constraint that the spin chain probe must either be antiferromagnetic (all $J > 0$) or ferromagnetic (all $J < 0$), while the signs of the local magnetic fields can be arbitrarily varied. In the following, we numerically optimize $H_{\text{XYZ}}$ and $H_{\text{XXX}}$ for global quantum thermometry. 

\subsection{Special case: XYZ probe}
As before, we minimize the average variance  $G (T_{\text{min}}, T_{\text{max}})$ for various $T_{\text{min}}$ and $T_{\text{max}}$ over the parameters in $H_{\text{XYZ}}$. Interestingly, the optimization yields the classical (longitudinal-field) Ising model with $J_x = J_y = h_x = h_y = 0$ (omitting the spin indices). Having the Ising probe as the optimal solution also means that the same results would apply for the Heisenberg XXZ model as well.
	\begin{figure*}
\subfloat{%
  \includegraphics[width=0\linewidth]{example-image-a}%
  \label{fig:ising4_bifurcation_J}
}\hfill
\subfloat{%
  \includegraphics[width=0\linewidth]{example-image-a}%
  \label{fig:ising4_bifurcation_h}
}\hfill
\subfloat{%
  \includegraphics[width=\linewidth]{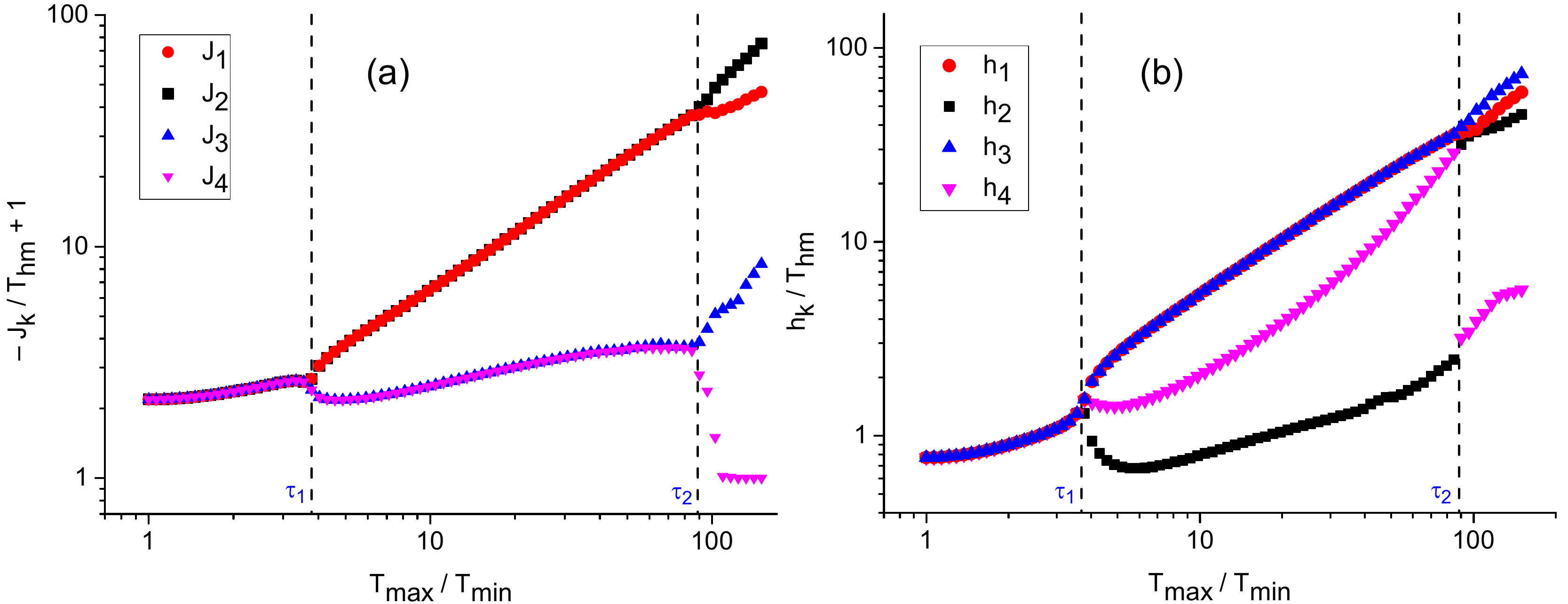}%
  \label{}
}\hfill
	\caption{\textbf{Optimal probe for $H_{\text{XYZ}}$.} Optimal parameters against temperature ratio $T_\text{max}/T_\text{min}$ for an optimized $H_{\text{XYZ}}$ (reduced to an Ising model) for $n = 4$ spins. (a) ZZ-coupling strengths $J_k$. For illustration purposes, we plot $-J_k$ shifted positively by 1 unit. (b) External magnetic fields. The parameters are scaled with the characteristic temperature $T_{hm}$, where $T_{hm}$ is the harmonic mean of $T_\text{min}$ and $T_\text{max}$. The dashed lines indicate the onset of the splitting of optimal parameters.}
	\label{fig:ising4_bifurcation}
	\end{figure*}
Figs. \subref*{fig:ising4_bifurcation_J} and \subref*{fig:ising4_bifurcation_h} shows the optimal Ising couplings and magnetic fields obtained from $H_{\text{XYZ}}$ for $n = 4$ spins. For thermometry in Phase 1, it turns out that the optimal solution is a uniform ferromagnetic Ising model with $J_z^{i} = -J < 0$ and $h_z^{i} = h > 0$. We note that the solution is not unique, and an antiferromagnetic Ising model with $J_z^{i} = J > 0$ is also possible with a different $h_z^{i}$. Interestingly, at the critical ratios indicated by the dashed lines, the uniformity of the optimal parameters breaks down, and the optimal parameters show a more sophisticated pattern. For more details, see Appendix \ref{appendixC}.

\subsection{Special case: XXX probe}

\begin{figure}
\subfloat{%
  \includegraphics[width=\linewidth]{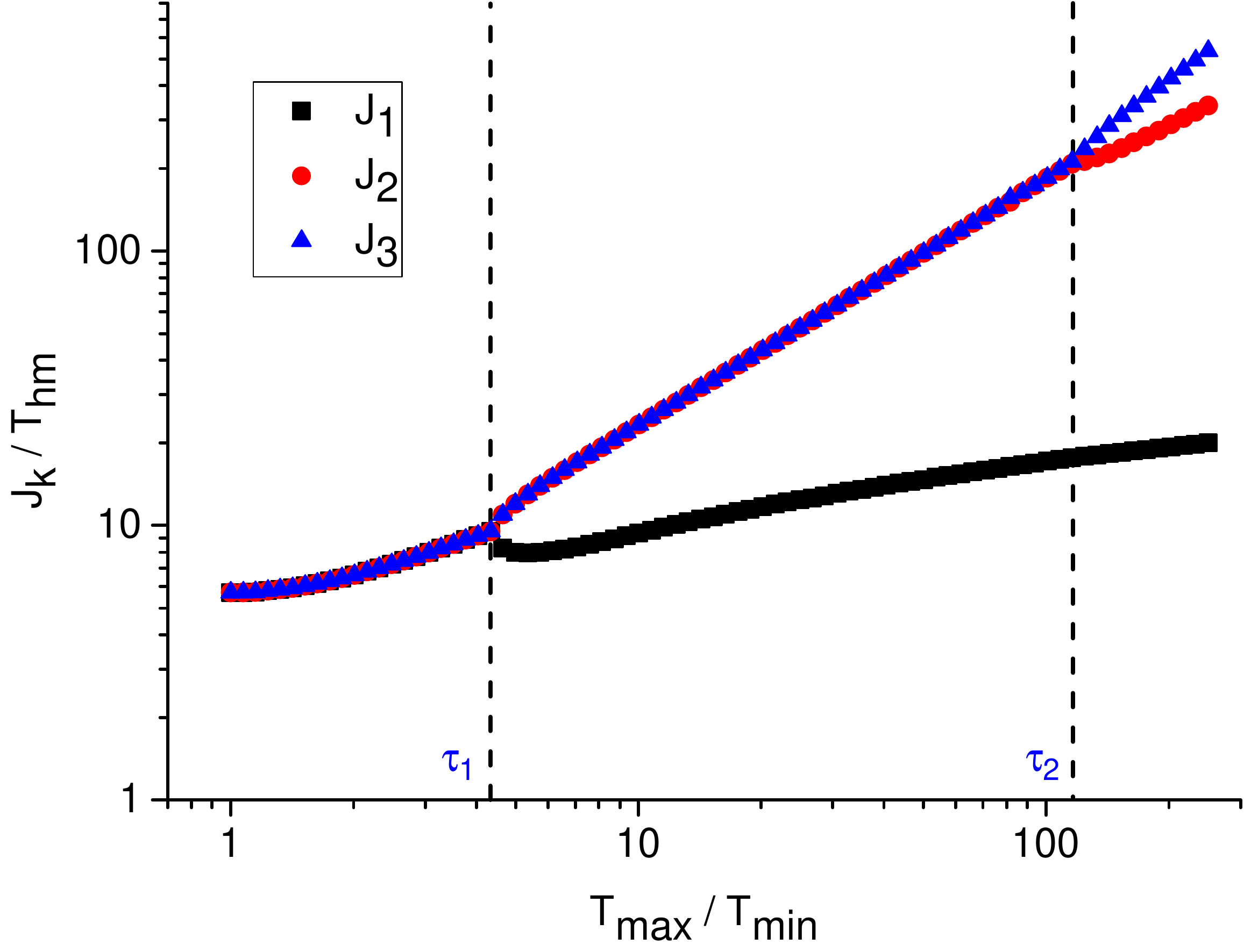}%
}\hfill
	\caption{\textbf{Optimal probe for $H_{\text{XXX}}$.} Optimal coupling parameters $J_k$ against temperature ratio $T_\text{max}/T_\text{min}$ for an optimized $H_{\text{XXX}}$ (reduced to an dimerized chain) for $n=6$ spins. The coupling parameters are scaled with the characteristic temperature $T_{hm}$, where $T_{hm}$ is the harmonic mean of $T_\text{min}$ and $T_\text{max}$. The dashed lines indicate the onset of the splitting of optimal parameters.}
	\label{fig:dimer3_bifurcation}
	\end{figure}

For the XXX probe, for the sake of simplicity, we first impose another constraint on $H_{\text{XXX}}$ by setting a homogenous magnetic field in each direction, i.e. $h_x^{i} = h_x, h_y^{i} = h_y, h_z^{i} = h_z \, \forall i$. Surprisingly, the optimal probe in this case turns out to be a fully dimerized spin chain with $J^{i} > 0$ for odd $i$ and $J^{i} = 0$ otherwise. The optimal solutions require no external magnetic field ($h_x = h_y = h_z = 0$). Fig. \ref{fig:dimer3_bifurcation} shows the optimal couplings for the dimer chain obtained from $H_{\text{XXX}}$, for $n = 6$ spins. This gives us three independent dimers with coupling parameters $J_1$, $J_2$ and $J_3$. In Phase 1, the dimers are all identical to one another. Again, at the critical ratios denoted by the dashed lines, the coupling parameters split into multiple branches.
 
One can physically explain why the fully dimerized spin chain is the optimal solution for $H_{\text{XXX}}$. The Heisenberg XXX Hamiltonian has a triply-degenerate first excited state for all even values of $n$. Having a uniform dimerized chain gives a $3n/2$-fold degeneracy in the first excited state, which is the maximum attainable value for the XXX Hamiltonian. That is why the fully dimerized chain becomes the optimal probe in Phase 1. As new phases emerge, one of the couplings of the dimers changes to vary the degeneracy of the spectrum, replicating the bifurcations found for the optimal probes. It is also noted that if we relax the constraint of homogeneous magnetic fields in $H_{\text{XXX}}$, we obtain a probe that is only marginally better than the dimerized chain. 

\subsection{Optimal measurements}

In general, energy measurements is very difficult in practice, making the saturation of Cram\'{e}r-Rao bound challenging. Remarkably, the probes obtained from optimizing $H_\text{XYZ}$ and $H_\text{XXX}$ are not just solutions that minimize ${G}$, but also have the added advantage of being easily measurable in the energy basis. In the case of $H_{\text{XYZ}}$ where the longitudinal-field Ising model is optimal, the energy measurements are equivalent to measuring all the spins in the computational basis, i.e. $\sigma_z$, which has been realized in various physical setups~\cite{PhysRevX.6.031007,Kokail2019}. In the case of $H_{\text{XXX}}$ where the fully dimerized chain is the optimal probe, the energy measurements is reduced to a series of singlet-triplet measurements on all the dimer pairs, which is a typical measurement in quantum dot arrays~\cite{Petta2180,Fogarty2018}.

Thus, the optimization results discussed earlier can saturate the Cram\'{e}r-Rao bound on top of having a low ${G}$. 

\subsection{Comparison with theoretical limit}

\begin{figure*}
	\centering
\subfloat{%
  \includegraphics[width=0\linewidth]{example-image-a}%
  \label{fig:localthermo_energies}%
}\hfill
\subfloat{%
  \includegraphics[width=0\linewidth]{example-image-a}%
  \label{fig:globalthermo_energies}%
}\hfill
\subfloat{%
  \includegraphics[width=0\linewidth]{example-image-a}%
  \label{fig:rel_diff_G}%
}\hfill
\subfloat{%
  \includegraphics[width=\linewidth]{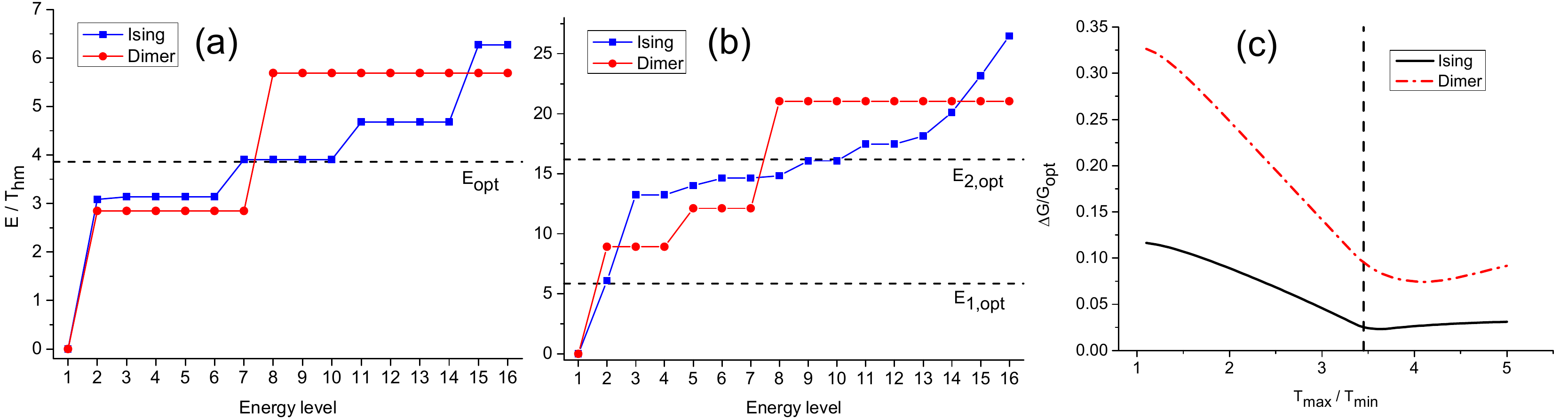}%
  \label{}%
}\hfill
	\caption{\textbf{Comparison of global thermometers.} Comparison of energy levels between the Ising model (blue squares, obtained from optimizing $H_{\text{XYZ}}$) and the dimerized chain (in red circles, obtained from optimizing $H_{\text{XXX}}$). The number of spins is $n = 4$ ($N=2^n=16$), and the optimal energies (dashed lines) are calculated two different temperature ratios: (a) $T_{\text{max}}/T_{\text{min}}=1$ (Phase 1). (b) $T_\text{max}/T_{\text{max}} = 5$ (Phase 2). The Ising energy levels are on average of closer proximity to the optimal energies. (c) Relative difference in $G$ against temperature ratio $T_\text{max}/T_\text{min}$ for the Ising model (in black, obtained from optimizing $H_{\text{XYZ}}$) and the dimerized chain (in red, obtained from optimizing $H_{\text{XXX}}$). The number of spins is $n = 4$, compared to the ideal case with $N = 16$ energy levels. The dashed line indicate the critical ratio for the ideal probe.}
	\label{fig:energy_compare}
	\end{figure*}

It is insightful to study the energy spectra of the optimized spin chain probes for $H_{\text{XYZ}}$ and $H_{\text{XXX}}$. For optimal probes of size $n = 4$, we plot in Figs. \subref*{fig:localthermo_energies} and \subref*{fig:globalthermo_energies} the energy spectra for $T_{\text{max}}/T_{\text{min}} = 1$ (Phase 1) and $T_{\text{max}}/T_{\text{min}} = 5$ (Phase 2), respectively. For Phase 1, the Ising excited energy levels shown in Fig. \subref*{fig:localthermo_energies} are on average closer to the optimal energy $E_\text{opt}$, denoted by the dashed line. Likewise, considering the case of Phase 2 thermometry in Fig. \subref*{fig:globalthermo_energies}, the Ising energy levels are closer to the respective optimal energies $E_\text{1,opt}$ (degeneracy = 1) and $E_{\text{2,opt}}$ (degeneracy = $N - 2$). In particular, we see that the Ising model has a single first excited state very close to $E_{\text{1,opt}}$, whereas the first excited state of the dimer must be 3-fold degenerate. This suggests that the Ising probe performs better than the dimerized probe in both cases. This is not surprising since $H_{\text{XYZ}}$ is more general than $H_{\text{XXX}}$, so we must always have $G_{\text{Ising}} \leq G_{\text{dimer}}$.

In order to quantify the quality of the optimal spin chain probes, one has to compare their performance with the theoretically ideal thermometers.
To do so, we compute the relative difference in ${G}$ with ${G}_{\text{opt}}$, i.e. $\Delta G/G_{\text{opt}} \equiv G/G_{\text{opt}} - 1$ for probes with $n = 4$, over a range of $T_{\text{max}}/T_{\text{min}}$ for which two phases are observable. When $\Delta G$ is small, it means that the probes perform closer to the theoretical limit. The results are plotted in Fig. \subref*{fig:rel_diff_G}. Remarkably, the Ising and dimerized probes perform closer to the theoretical limit in Phase 2 as compared to Phase 1. In addition, as expected, the Ising probe indeed outperforms the dimerized probe at all values of $T_{\text{max}}/T_{\text{min}}$.


\section{Numerical methods}
\label{sec:methods}

The numerical optimization in this work is done primarily using global optimization routines. In particular, we use the SciPy implementation ~\cite{2020SciPy-NMeth} of differential evolution~\cite{diffev}, which is an example of an evolutionary algorithm. Differential evolution is a metaheuristic search algorithm that iteratively improves the candidate solution by evolving the population which navigates the search space. It does not rely on gradient information, making it suitable for exploring large search spaces which is useful for optimizing over large number of energy levels. In all cases, we use a population of 500, and increasing the population further does not improve the quality of the optimal solution found. The results are also verified with other popular global optimization routines such as simulated annealing and basin-hopping. Hence, our results do not depend on the specific optimization algorithm used.

For spin chains with $n \geq 10$, global optimization becomes infeasible due to two sources of computational costs: (i) diagonalization of the Hamiltonian (in an exponentially large Hilbert space) to obtain all the eigenenergies at each step, and (ii) large number of iterations required for global optimization. While it is possible to speed up the computation by restricting the diagonalization to only return a subset of the lowest eigenenergies, we found that the optimal results are rather inaccurate unless most of the eigenvalues are retained. Hence, the speed up from this approach is marginal. In order to proceed with the optimization for larger systems, we employ a transfer learning approach where the optimal probe for a smaller $N$ is recursively used as an initial guess for an incrementally larger system size, and perform local optimization using the gradient-based BFGS algorithm (see Appendix \ref{appendixC}). 

\section{Conclusion}

In this paper, we quantitatively formalize the notion of global thermometry, which operates over arbitrary large temperature ranges. This is achieved by minimizing the average of variance over a given temperature interval. The proposed approach naturally provides the energy structure of an optimal probe. Interestingly, depending on the temperature range, the optimal thermometer goes through different phases, each favouring a particular energy structure. We provide a realization of such optimal probes in generalized Ising models with multi-particle interactions. In practice, however, such interactions might be challenging to fabricate. Hence, by imposing further constraints, we provide new designs for optimal probes which are limited to two-body interactions in the form of nearest neighbours of special type of interactions. Starting from a non-uniform XYZ Hamiltonians, our optimization results in a uniform Ising model as the optimal thermometer which can be realized in ion traps. On the other hand, if we restrict ourselves to non-uniform Heisenberg models, then the optimal thermometer is a dimerized chain which can be realized in quantum dot arrays. We also observe the appearance of various phases for practically optimal probes as we vary the temperature range.

Our results unlock several fresh research avenues. For instance, how one can generalize the developed global thermometry procedure to non-equilibrium probes. Another direction to pursue is to find optimal designs for thermometers in other physical platforms, such as optomechanical systems or itinerant particles. \\

\begin{acknowledgments}
AB acknowledges the National Key R\&D Program
of China, Grant No. 2018YFA0306703. KB and LCK are grateful to the National Research Foundation and the Ministry of Education, Singapore for financial support. The authors thank Jingu Pang for the drawing of the schematic diagram.

\end{acknowledgments}
\bibliographystyle{apsrev4-1}
\bibliography{thermo_bib}

\onecolumngrid
\appendix
\newpage
\counterwithin{figure}{section}
\section{Optimal probe spectrum in the narrow-range regime}
\label{appendixA}
In the narrow-range regime where $T_{\text{max}}/T_{\text{min}}$ is near unity, the optimal conditions can be evaluated. From numerical simulations in Fig. \ref{fig:qudit_bifurcation}, we know that the optimal probe in Phase $1$ thermometry is an effective two-level system with (N-1)-fold degeneracy in the excited state. Using this Ansatz, the quantum Fisher information is (setting $k_B = 1$):
    \begin{equation}
\mathcal{F}_\text{th} (\epsilon, N, T) = \frac{(N-1)\epsilon^2 e^{\epsilon/T}}{(N-1+e^{\epsilon/T})^2 T^4}
    \end{equation}
where $\epsilon$ is the energy gap between the ground and excited state. The measure for global thermometry in Eq. (\ref{eq:global_measure}) is proportional to the integral of $1/\mathcal{F}_{\text{th}}$ against temperature in the range $[T_0, T_0 + \delta]$:
    \begin{equation}
g = \int_{T_0}^{T_0 + \delta} \frac{dT}{\mathcal{F}_{\text{th}}} = \frac{e^{-\epsilon/T_0} (N-1 + e^{\epsilon/T_0})^2 T_0^4 \delta}{(N-1) \epsilon^2} + \frac{e^{-\epsilon/T_0} T_0^2 \delta [ 4(N-1 + e^{\epsilon/T_0})^2 T_0 \delta + [(N-1)^2 + e^{-2\epsilon/T_0}] \delta \epsilon ] }{2(N-1)\epsilon^2} + \mathcal{O}{(\delta^3)}
    \end{equation}
where we have assumed $\delta \ll T_0$. To minimize the integral, we define dimensionless variables $x \equiv \epsilon/T_0$ and $\delta^\prime \equiv \delta/T_0$ and set the partial derivative $\partial g / \partial \epsilon$ to zero. We thus obtain the optimal condition
    \begin{equation}
8(N-1)e^x (1+2 \delta^\prime) + e^{2x} [4(1+2\delta^\prime) - (2+5\delta^\prime)x + \delta^\prime x^2] + (N-1)^2 [4 + 5\delta^\prime x + \delta^\prime x^2 + 2(4\delta^\prime + x)] = 0
    \label{eq:global_optimal_condition}
    \end{equation}
which can be solved numerically to get the optimal energy gap. To get further insights, we can solve for $e^x$ implicitly from Eq. (\ref{eq:global_optimal_condition}) which is quadratic in $e^x$, and expand the solution in powers of $\delta^\prime$. This gives
    \begin{equation}
e^x = (N-1) \frac{x+2}{x-2} + (N-1) \frac{x^3 \delta^{\prime}}{2(x-2)^2} + (N-1) \frac{x^3 [12 + x(x-6)] \delta^{\prime 2}}{(x-2)^3} + \mathcal{O}(\delta^{\prime 3})
    \label{eq:corrections}
    \end{equation}
Note that the zeroth-order term exactly recovers the optimal condition in Eq. (\ref{eq:local_opt}) discovered in Ref. \cite{Louis2015}. Thus, we have found finite-range corrections to the optimal probe for local thermometry. Note that if we expanded $g$ to only first order, we do not obtain the correction terms. As an example, we set $T_0 = 1$ and $\delta = 0.1$. Numerical optimization gives the optimal energy gap $x \equiv \epsilon/T = 4.05268$, while the second-order correction in Eq. (\ref{eq:corrections}) give $x = 4.04085$, which is only $0.292\%$ away from the numerical result.

\section{Generalized Ising model as a universal classical Hamiltonian simulator}
\label{appendixB}

In the main text, we pointed out that the generalized Ising model comprising all $k$-body ZZ-interactions with $k \leq n$ can realize the optimal energy spectra for global thermometry. We provide an intuitive justification based on numerical evidence, leaving a rigorous investigation for future studies. To obtain the optimal two-level solution, one needs to control $2^n - 1$ energy levels using the Hamiltonian parameters, in order to engineer the maximal degeneracy (the remaining level can be chosen as the ground state without loss of generality). In this generalized Ising model, we have exactly $2^n - 1$ free parameters $\{ J_k \}$. Moreover, since the Hamiltonian is diagonal, the eigenenergies are simply a linear combination of the $J_k$ parameters. As a result, all the energy levels can be controlled independently, and the degeneracy condition can be easily achieved. 

As an example, let us consider the $n = 4$ case. This gives us 4 $J_1$ parameters, 6 $J_2$ parameters, 4 $J_3$ parameters and 1 $J_4$ parameter for a total of 15 free parameters. Setting the magnitude of all the parameters to be the same, and choosing the signs (for instance) to be $J_1 : (- - + +)$, $J_2: (- + + + + -)$, $J_3: (+ + - -)$, $J_4: (-)$, the optimal solution can be obtained with energy gap $2^n J$. 

Following a similar procedure, one can also fix the magnitude of $2^{(n-1)}$ parameters to be some $J$ and the rest of the parameters to have magnitude $K$, with $J < K$. Choosing appropriate signs, the energy spectrum with 1 ground state (with zero energy), 1 first excited state (with energy $2^n J$) and $(2^n - 2)$ degenerate second excited states (with energy $2^{n-1} (J+K)$ ) can be obtained, which is the optimal solution for Phase 2 thermometry.

Based on strong numerical evidence, we posit that the generalized Ising model can simulate any energy spectrum. To see this, we can write down the eigenvalues of the generalized Hamiltonian in Eq. (\ref{eq:generalized_ising}) as a $2^n \times 2^n-1$ coefficient matrix $\mathcal{A}$ with the rows denoting the Ising eigenstates and the columns denoting the various $J_k$ parameters. Since the eigenvalues are linear in $J_k$, finding the coupling parameters $J_k$ reduces to a linear algebra problem $\mathcal{A} \bold{J} = \bold{E}$, where $\bold{J}$ is the column vector containing the $2^n - 1$ parameters $J_k$, and $\bold{E}$ is the column vector containing the $2^n$ eigenenergies. The rank of $\mathcal{A}$ is upper bounded by $2^n - 1$, so we are free to set any one row in $\mathcal{A}$ to zero. It can be checked that the eigenenergy of $\ket{1 1 1 \ldots 1}$ is $(-1) \times$ (the sum of all other eigenenergies). Thus, without loss of generality, we can truncate the row corresponding to $\ket{1 1 1 \ldots 1}$ in $\mathcal{A}$, and set the corresponding eigenenergy to zero (which can be truncated). Thus, $\mathcal{A}$ is now a square matrix of dimension $2^n - 1$. For a unique solution of $\mathcal{A} \bold{J} = \bold{E}$ to exist, the rank of $\mathcal{A}$ must be $2^n - 1$. While this remains to be proven, we provide some justification for why $\mathcal{A}$ might be full rank: Suppose the contrary that one of the rows in $\mathcal{A}$ can be written as a linear combination of $k$ other rows, where $k$ is arbitrary. Particular to the Ising model, we know that the coefficient of $J_n$ is a product of the coefficients in $J_1$. Writing the $J_1$ coefficients as $x_{ij}$, with $i = 1,2, \ldots, k$ and $j = 1,2,\ldots,n$, the linear combination of the $J_1$ coefficients is thus $\sum_{i=1}^{k} c_i x_{ij}$. If we first construct the $J_n$ coefficients via the product $\prod_{j=1}^{n} x_{ij}$ and perform the linear combination of the $J_n$ coefficients, we obtain the resultant $J_n$ coefficient $\sum_{i=1}^{k} \prod_{j=1}^n c_i x_{ij}$. However, if we first perform the linear combination of the $J_1$ coefficients $\sum_{i=1}^{k} c_i x_{ij}$ and construct the resultant $J_n$ via the product, we instead obtain $\prod_{j=1}^n \sum_{i=1}^k c_i x_{ij}$. For \text{rank}($\mathcal{A}$) $\neq$ $2^n - 1$, it would require that $\sum_{i=1}^{k} \prod_{j=1}^n c_i x_{ij} = \prod_{j=1}^n \sum_{i=1}^k c_i x_{ij}$ (i.e. the sum commutes with the product). To reinforce our claim, we conducted numerical tests for $\mathcal{A}$ up to $n=10$ spins, and, we found the rank to be $2^n-1$.

\section{Optimization results for longer spin chains}
\label{appendixC}

\begin{figure}
\subfloat{%
  \includegraphics[width=0\linewidth]{example-image-a}%
  \label{fig:ising_transfer}
}\hfill
\subfloat{%
  \includegraphics[width=0\linewidth]{example-image-a}%
  \label{fig:XYZ13_noise}
}\hfill
\subfloat{%
  \includegraphics[width=\linewidth]{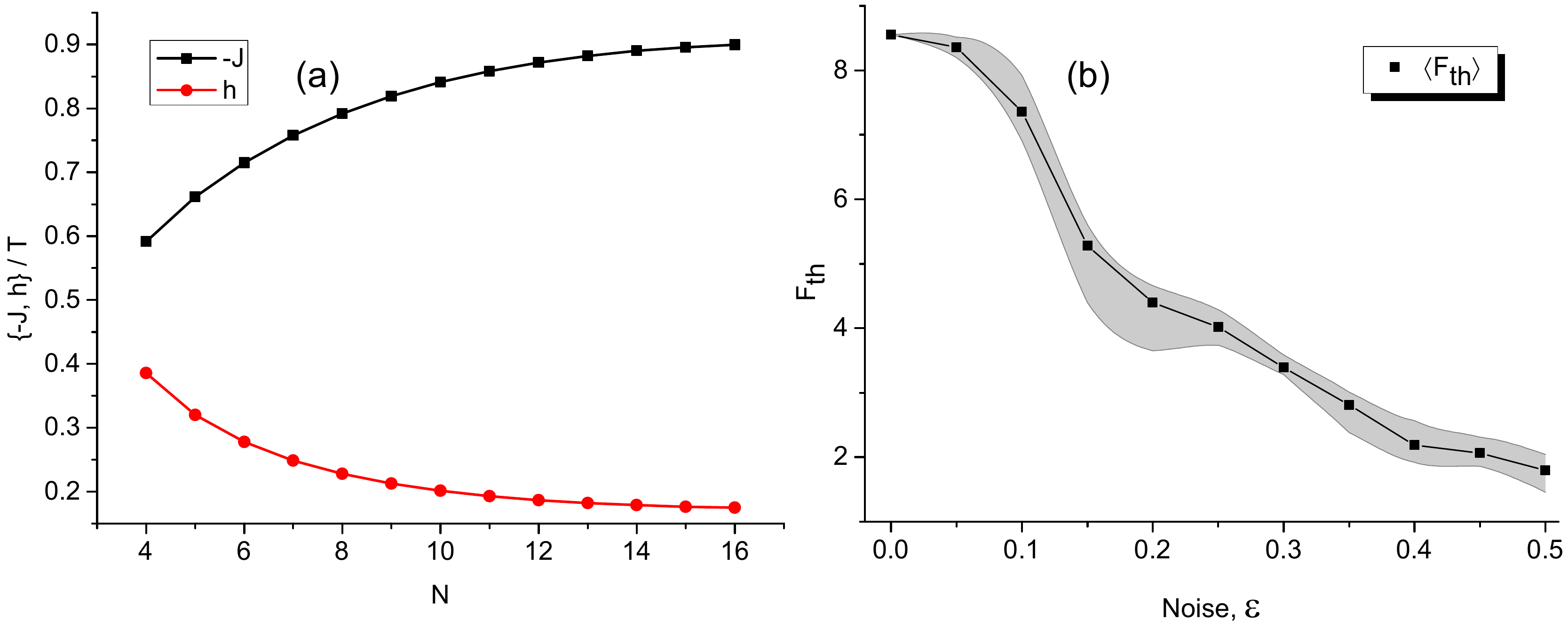}%
}\hfill
	\caption{(a) Optimal probe for $H_{\text{XYZ}}$ for $4 \leq n \leq 16$, using the transfer learning method. The optimal solution found is the longitudinal-field Ising model with uniform coupling and magnetic field. (b) Addition of a random noise in $H_{\text{XYZ}}$ for $n = 13$. For each value of $\epsilon$, the quantum Fisher information $\mathcal{F}_{\text{th}}$ is averaged over 30 runs. The average value is plotted, with the gray band denoting the range of values found.}
	\label{}
	\end{figure}

As discussed in Sec. \ref{sec:methods}, when the number of spins, $n$, is large, the Hilbert space dimension grows exponentially. As such, the diagonalization becomes exponentially difficult. This, combined with the large number of iterations required to reach a global optimum, renders the direct global optimization method infeasible for large $n$. One solution to get around this problem is to use a transfer learning approach where the optimal solution for a small $n$ is modified and used as the ansatz for an incrementally larger $n$, which hopefully speeds up the convergence to an optimal solution. Here, we demonstrate this approach for $H_{\text{XYZ}}$ for $n$ up to 16. For simplicity, we consider the regime of local thermometry which is a subset of Phase 1 thermometry. The optimization yields the longitudinal-field Ising model with a uniform coupling $J < 0$ and uniform magnetic field $h$, as shown in Fig. \subref*{fig:ising_transfer}. Of course, this approach does not guarantee a globally optimal solution. To check that the solution found is at least a local optimum, we can use the optimal parameters in the initial Hamiltonian and add an independent random fluctuation drawn from a uniform distribution in $[-\epsilon, \epsilon]$ ($\epsilon$ is the noise level) to all the free Hamiltonian parameters. If the solution is indeed optimal, then the addition of any noise will lower the performance of the probe. We show this for the case of $n = 13$ in Fig. \subref*{fig:XYZ13_noise}. For each noise level $\epsilon$, we calculate the quantum Fisher information $\mathcal{F}_{\text{th}}$ of the probe (with noise), averaged over 30 independent runs. Evidently, the addition of noise worsens the probe performance compared to the noiseless case.

\end{document}